\documentclass[secnumarabic,amssymb, nobibnotes, aps, prb, twocolumn]{revtex4}

\usepackage{color}
\usepackage{epsfig, graphics, graphicx, subfigure, wrapfig, lipsum, amsmath}
\usepackage{verbatim}
\usepackage{lipsum} 
\usepackage{dcolumn}
\usepackage{bm}
\usepackage{amsmath, braket}
\usepackage{array}
\usepackage{xr}

\newcommand{\roml}[1]{\lowercase\expandafter{\romannumeral #1\relax}}
\newcommand{\romu}[1]{\uppercase\expandafter{\romannumeral #1\relax}}

\begin{document}

\title{Anharmonic lattice dynamics and thermal transport in type-I inorganic clathrates}

\author{Shravan Godse}
\author{Yagyank Srivastava}
\author{Ankit Jain}
\email{a\_jain@iitb.ac.in}
\affiliation{Mechanical Engineering Department, IIT Bombay, India}
\date{\today}%

\begin{abstract}
The anharmonic phonon properties of type-I filled inorganic clathrates Ba$_{8}$Ga$_{16}$Ge$_{30}$ and Sr$_{8}$Ga$_{16}$Ge$_{30}$ are obtained from the first-principles calculations by considering the temperature-dependent sampling of the potential energy surface and quartic phonon renormalization.  Owing to the weak binding of guest atoms with the host lattice, the obtained guest modes  undergo strong renormalization with temperature and become stiffer by up to 50\% at room temperature in Sr$_{8}$Ga$_{16}$Ge$_{30}$. The calculated phonon frequencies and associated thermal mean squared displacements are comparable with experiments despite the on-centering of guest atoms at cage centers in both clathrates. 
{\color{black} Lattice thermal conductivities are obtained in the temperature range of 50-300 K accounting for three-phonon scattering processes and multi-channel thermal transport. The contribution of coherent transport channel is significant at room temperature (13\% and 22\% in Ba$_{8}$Ga$_{16}$Ge$_{30}$ and Sr$_{8}$Ga$_{16}$Ge$_{30}$) but is insufficient to explain the experimentally observed glass-like thermal transport in  Sr$_{8}$Ga$_{16}$Ge$_{30}$.}
\end{abstract}

\maketitle

\section{Introduction}

\begin{figure}
\begin{center}
\includegraphics{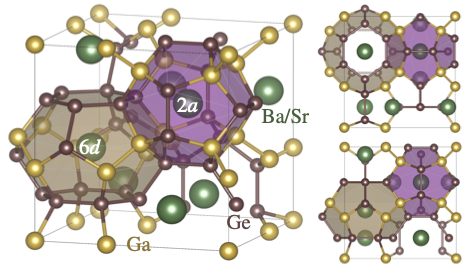}
\end{center}
\caption{Crystal structure of type-I intermetallic clathrate. The structure is composed of two types of polyhedral cages: 20 atoms-pentagonal dodecahedra cages (highlighted in magenta color) and 24 atoms-tetrakaidecahedra cages (brown color). These cages host 2$a$ and 6$d$ Wyckoff sites and are present in the ratio of 2:6 in crystal structure.}
\label{fig_structure}
\end{figure}

Type-I intermetallic clathrates, such as Ba$_{8}$Ga$_{16}$Ge$_{30}$, Sr$_{8}$Ga$_{16}$Ge$_{30}$ have an open cage-like crystal structure (Fig.~\ref{fig_structure}) and are actively investigated for potential application in thermoelectric energy generation due to their phonon-glass and electron-crystal like properties \cite{slack1995crc,Takabatake_2014}. The cages in these clathrates can be occupied by heavy guest atoms (referred to as rattlers) which are loosely bound to the host lattice.  The rattlers have large thermal amplitudes around their equilibrium positions producing several interesting effects like the peculiar lattice dynamics which result in flat low-frequency phonon modes with avoided crossings and low lattice thermal conductivity\cite{Cohn1999,SALES1999528,Chakaumakos2000,keppens2000,Hermann2005,Baumbach2005,chulholee2008}. 

The thermal properties of type-I ternary clathrates X$_8$Ga$_{16}$Ge$_{30}$ (X: Ba, Sr, henceforth referred to as BGG and SGG respectively) have attracted many experimental and computational research studies \cite{sales2001,Bentien2004,Tadano2018}.
Interestingly,  while BGG behaves like a typical crystalline material and has a decreasing thermal conductivity with rising temperature, SGG shows a glasslike behaviour with reverse temperature-dependence \cite{sales2001}.
The measured thermal displacement of rattlers in SGG are abnormally large and inspired the development of off-centered multi-positioned rattler theory\cite{Nakayama2011}. The off-centering of rattlers is proposed as the  reason  for  glass-like  thermal  conductivity of  SGG,  though  similar  glass-like  thermal  conductivity  is  also reported for other clathrates without off-centering of rattlers {\color{black} with p- vs n-type doping of the material\cite{Avila2006, Bentein2006}}. Further, {\color{black} depending on the growth conditions}, crystal-like thermal conductivity is measured now for several clathrates with potential off-centering of rattlers\cite{Christensen2016}.

The computations are not helpful  and for SGG, the phonon dispersion obtained from ab-initio driven finite-difference approach either resulted in imaginary phonon modes or the  mode-frequencies obtained for low-lying optical phonons are severe under-prediction of the experimentally measured values\cite{Madsen2005, Christensen2016}. 
The large thermal displacements of rattlers render strongly temperature-dependent   force interactions, not accounted for by lowest-order theories considered so far in literature \cite{Christensen2006A,Christensen2009A}.

Here, using temperature-dependent force constants and phonon renormalization theory\cite{Jain2020}, the anharmonic phonon properties of BGG and SGG clathrates are obtained in the temperature range of 50-300 K. {\color{black}We report the phonon frequencies, thermal displacements, and three-phonon scattering rates in the considered temperature range. Further, using the multichannel thermal transport theory, we also report  the temperature-dependent contribution to thermal transport from particle-like phonon and wave-like coherent transport channels\cite{simoncelli2019}.
}

\section{Theory and Computational Details} 
\label{sec_theory_comp}

The harmonic phonon frequencies, $\omega_{{\boldsymbol{q}}\nu}$, and eigenvcetors, ${\boldsymbol{e}}_{{\boldsymbol{q}}\nu}$, are obtained by diagonalizing the dynamical matrix, ${\boldsymbol{D}}_{{\boldsymbol{q}}}$, obtained as \cite{reissland1973, dove1993}:
\begin{equation}
 \label{eqn_theory_dynamical}
 D_{\boldsymbol{q}}^ {3b+\alpha, 3b^{'} + \beta} = \frac{1}{\sqrt{m_b m_{b^{'}}}} \sum_{l^{'}} \Phi_{b0;b^{'}l^{'}}^{\alpha\beta} 
 e^{i{\boldsymbol{q}}.( {\boldsymbol{r}}_{b^{'}l^{'}}  -  {\boldsymbol{r}}_{b0}  ) },
\end{equation}
where the summation is over all unit-cells in the lattice (N), $m_b$ is the mass of atom $b$ in the unit-cell,  $\boldsymbol{r}_{bl}$ is the position vector of atom $b$ in the $l^{th}$ unit-cell,  $\Phi_{ij}^{\alpha\beta}$ is the real-space ($ij, \alpha\beta$)-element of the  harmonic force constant matrix $\boldsymbol{\Phi}$, and ${\boldsymbol{q}}$, $\nu$ are phonon wavevector and polarization. The obtained phonon frequencies (and eigenvectors) are self-consistently corrected for higher-order anharmonicity via renormalization which requires higher-order anharmonic force constants.

While the anharmonic force constants can be obtained from the finite-difference of density functional theory (DFT) forces by displacing one or more atoms in the computational supercell corresponding to the equilibrium positions of atoms, the force constants obtained from this approach samples the PES at 0 K. For moderately anharmonic solids, this 0 K sampling is adequate and results in a minimal error. For strongly anharmonic solids, such as clathrates, the PES anharmonicity is strongly dependent on atomic displacements and is a strong function of temperature. In this work, T-dependent force constants are obtained by force-displacement data fitting on thermally populated supercells. 
The thermal displacements, $u_{b,l}^{\alpha}$, of atoms in computational cells are  obtained as\cite{west2006},
\begin{equation}
    \begin{split}
        \label{eqn_thermal_population}
        u_{b,l}^{\alpha} = \frac{1}{\sqrt{N}}
        \sum_{\boldsymbol{q}\nu}
        \sqrt{\frac{\hbar(n_{\boldsymbol{q}\nu} + 1)}
            {m_b\omega_{\boldsymbol{q}\nu}}}
            \sin{(2\pi\eta_{1, \boldsymbol{q}\nu})}
            \\
            \sqrt{-\ln{(1-\eta_{2, \boldsymbol{q}\nu})}}
            {{\tilde{e}^\alpha}_{b, \boldsymbol{q}\nu}}
            e^{i \boldsymbol{q}  \cdot{\boldsymbol{r}}_{0l} },
    \end{split}
\end{equation}
where $\eta_{1, \boldsymbol{q}\nu}$ and $\eta_{2, \boldsymbol{q}\nu}$ are random numbers sampled from a uniform distribution and constrained by $\eta_{1, \boldsymbol{q}\nu} = \eta_{1, -\boldsymbol{q}\nu}$ and $\eta_{2, \boldsymbol{q}\nu} = \eta_{2, -\boldsymbol{q}\nu}$. The forces corresponding to these thermally displaced snapshots are obtained from DFT calculations. The temperature is taken into account in Eqn.~\ref{eqn_thermal_population} via T-dependent mode population ($n_{\boldsymbol{q}\nu}$) and the 
phonon frequencies are obtained from Eqn.~\ref{eqn_theory_dynamical} via diagonalization of the bare harmonic force constants [as obtained from the density functional perturbation theory (DFPT)] based dynamical matrix.

{\color{black}
While extracting anharmonic force constants via  force-displacement data fitting, the contribution of the corrected harmonic force constants (see Eqn.~\ref{eqn_renorm} below) is removed from DFT forces. The anharmonic force constants are obtained using Taylor-series fitting to residual-force displacement dataset.  The corrected harmonic force constants,  $\Phi^{c, \alpha\beta}_{ij}$, are obtained via self-consistent renormalization as\cite{wallace1972}: 
\begin{eqnarray}
\begin{split}
\Phi^{c, \alpha\beta}_{ij} = 
     \Phi^{o, \alpha\beta}_{ij} + 
    \frac{\hbar}{4N}\sum_{l^{'''}l^{''''}} \sum_{b^{'''}b^{''''}}\sum_{\gamma\delta}\sum_{\boldsymbol{q}\nu}
    \Xi_{ijkl}^{\alpha\beta\gamma\delta}
    \\
    \frac{{{\tilde{e}^\gamma}_{b^{'''}, \boldsymbol{q}\nu}}
    {{\tilde{e}^{\dagger\delta}}_{b^{''''}, \boldsymbol{q}\nu}}}
    {\omega_{\boldsymbol{q}\nu}\sqrt{m_{b^{'''}} m_{b^{''''}}}}
    (2n_{\boldsymbol{q}\nu}+1)
    e^{i\boldsymbol{q}\cdot{ (\boldsymbol{r}_{0l^{'''}} - \boldsymbol{r}_{0l^{''''}}) }}, 
    \label{eqn_renorm}
\end{split}
\end{eqnarray}
where $\Phi^{o, \alpha\beta}_{ij}$ represent the bare harmonic force constants (as obtained from DFPT), and $n_{\boldsymbol{q}\nu}$, $\Xi_{ijkl}^{\alpha\beta\gamma\delta}$ are the Bose-Einstein distribution and quartic force constants (as obtained from residual-force displacement data fitting above). The corrected harmonic force constants are obtained by iterating through Eqn.~\ref{eqn_renorm} based on updated phonon frequencies and eigenvectors. Finally, the entire cycle of residual-force displacement data fitting and self-consistent renormalization is repeated until convergence to obtain the self-consistent set of corrected force constants. }

{\color{black} 
The phonon scattering rates are obtained by considering the three-phonon scattering processes as\cite{reissland1973, wallace1972, Jain2020}:

\begin{equation}
\begin{split}
 \label{eqn_rta_3ph}
 \frac{1}{\tau_{\boldsymbol{q}\nu}^{o,3ph}} 
 =
  \sum_{{\boldsymbol{q}_{1}}\nu_{1}}
 \sum_{{\boldsymbol{q}_{2}}\nu_{2}}
 \bigg\{
 \Big\{
  {(n_{{\boldsymbol{q}_{1}\nu_{1}}}  - n_{{\boldsymbol{q}_{2}\nu_{2}}})}
 W^{+}
  \Big\} 
  +
  \\
  \frac{1}{2}
   \Big\{
  (n_{{\boldsymbol{q}_{1}\nu_{1}}} + n_{{\boldsymbol{q}_{2}\nu_{2}}} +  1)
  W^{-}
   \Big\}
   \bigg\},
 \end{split}
 \end{equation}
where $\boldsymbol{W}$ represents scattering probability matrix given by:
\begin{equation}
    \begin{split}
    \label{eqn_W_3ph}
W^{\pm}       
=
\frac{2\pi}{\hbar^2}
 \left|
 \Psi_{ {\boldsymbol{q}} (\pm{\boldsymbol{q}_{1}}) (-{\boldsymbol{q}_{2}}) }^{\nu \nu_{1} \nu_{2}}
 \right|^2
  \delta({\omega_{{{\boldsymbol{q}}_{}}\nu_{}} \pm \omega_{{{\boldsymbol{q}}_{1}}\nu_{1}} - \omega_{{{\boldsymbol{q}}_{2}}\nu_{2}}}).
    \end{split}
\end{equation}
The $\Psi_{ {\boldsymbol{q}} {\boldsymbol{q}_{1}} {\boldsymbol{q}_{2}} }^{\nu \nu_{1} \nu_{2}}$ are the Fourier transform of real-space cubic constants, $\Psi^{\alpha\beta\gamma}_{bl;b^{'} l^{'};b^{''} l^{''}}$, and are obtained as:
\begin{equation}
    \begin{split}
        \label{eqn_cubic_IFC}
\Psi_{ {\boldsymbol{q}} {\boldsymbol{q}_{1}} {\boldsymbol{q}_{2}} }^{\nu \nu_{1} \nu_{2}}
=
 \Psi_{ {\boldsymbol{q}} {\boldsymbol{q}^{'}} {\boldsymbol{q}^{''}} }^{\nu \nu^{'} \nu^{''}} = 
 N
 {\left(\frac{\hbar}{2N}\right)}^{\frac{3}{2}}
 \sum_{b} \sum_{b^{'} l^{'}}
\sum_{b^{''} l^{''}} 
\sum_{\alpha\beta\gamma} 
\Psi^{\alpha\beta\gamma}_{bl;b^{'} l^{'};b^{''} l^{''}}
\\
\times
\frac{
{{\tilde{e}}_{b,\boldsymbol{q}\nu}^{\alpha}}  
{{\tilde{e}}_{b^{'},{\boldsymbol{q}}^{'} \nu^{'}}^{\beta}} 
{{\tilde{e}}_{b^{''},{\boldsymbol{q}}^{''} \nu^{''}}^{\gamma}} }{\sqrt{ 
{m_b \omega_{\boldsymbol{q}\nu}}  
{m_{b^{'}} \omega_{{\boldsymbol{q}}^{'}\nu^{'}}}   
{m_{b^{''}} \omega_{{\boldsymbol{q}}^{''}\nu^{''}}}   }}  
e^{[i( {{\boldsymbol{q}}^{'}}  \cdot{\boldsymbol{r}}_{0l^{'}} 
+    {{\boldsymbol{q}}^{''}}  \cdot{\boldsymbol{r}}_{0l^{''}} )]} ,
    \end{split}
\end{equation}
The $\delta$ in Eqn.~\ref{eqn_W_3ph} represents the delta-function ensuring energy conservation and the summation in Eqn.~\ref{eqn_cubic_IFC} is performed over phonon wavevectors satisfying crystal momentum conservation, i.e., $\boldsymbol{q} + \boldsymbol{q_1} + \boldsymbol{q_2} = \boldsymbol{G}$, where $\boldsymbol{G}$ is the reciprocal space lattice vector.

The total thermal conductivity is obtained as:
\begin{equation}
k^{tot} = k^{p} + k^{c},
\end{equation}
where $k^{p}$ and $k^{c}$ are the contributions of particle and coherent channel towards the thermal transport \cite{simoncelli2019}. 
The contribution of particle channel towards the thermal conductivity in the $\alpha$-direction is obtained by solving the Boltzmann transport equation and using the Fourier's law as \cite{reissland1973}:
\begin{equation}
 \label{eqn_theory_conduct}
 k_{\alpha}^p = \sum_{{\boldsymbol{q}}} \sum_{\nu} c_{{\boldsymbol{q}}\nu} v_{{\boldsymbol{q}}\nu, \alpha}^{2} \tau_{{\boldsymbol{q}}\nu, \alpha},
\end{equation}
where  $c_{{\boldsymbol{q}}\nu}$ is the phonon specific heat and $v_{{\boldsymbol{q}}\nu,\alpha}$ is the $\alpha$ component of phonon group velocity vector ${\boldsymbol{v}_{{\boldsymbol{q}}\nu}}$ (${\boldsymbol{v}_{{\boldsymbol{q}}\nu}} = \frac{\partial \omega_{\boldsymbol{q}\nu}}{\partial \boldsymbol{q}}$). The phonon specific heat are obtained from the phonon vibrational frequencies as:
\begin{equation}
\label{eqn_theory_cph}
c_{{\boldsymbol{q}}\nu} = \frac{\hbar\omega_{{\boldsymbol{q}}\nu}}{V} \frac{\partial n^{o}_{{\boldsymbol{q}}\nu}}{\partial T} = \frac{k_\text{B} x^2 e^x }{(e^x-1)^2},
\end{equation}
where $V$ is the crystal volume,  $k_\text{B}$ is the Boltzmann constant, and $x = \frac{\hbar\omega_{{\boldsymbol{q}}\nu}}{k_\text{B}T}$. 
The coherent contribution towards the lattice thermal conductivity is obtained using the formulation of Simoncelli et al.\cite{simoncelli2019} as:
\begin{equation}
\begin{split}
    \label{eqn_coherent}
k^c_{\alpha\beta} 
 = 
 \frac{\hbar^2}{k_BT^2}
 \frac{1}{VN}
 \sum_{\boldsymbol{q}}
 \sum_{(\nu_{} \ne \nu_{1})}
 \frac{ \omega_{\boldsymbol{q}\nu_{}} +  \omega_{\boldsymbol{q}\nu_{1}} }{2}
 V^{\alpha}_{\boldsymbol{q}, \nu\nu_{1}}
 V^{\beta}_{\boldsymbol{q}, \nu_1\nu_{}} 
 \times
 \\
 \frac{
  \omega_{\boldsymbol{q}\nu_{}}  n_{\boldsymbol{q}\nu_{}} (n_{\boldsymbol{q}\nu_{}} + 1)
  +
  \omega_{\boldsymbol{q}\nu_{1}}  n_{\boldsymbol{q}\nu_{1}} (n_{\boldsymbol{q}\nu_{1}} + 1)
 }
 {
 4(\omega_{\boldsymbol{q}\nu_{}} - \omega_{\boldsymbol{q}\nu_1})^2 
 + (\Gamma_{\boldsymbol{q}\nu_{}} + \Gamma_{\boldsymbol{q}\nu_{1}} )^2
 }
 (\Gamma_{\boldsymbol{q}\nu_{}} + \Gamma_{\boldsymbol{q}\nu_{1}}).
\end{split}
\end{equation}
$\Gamma_{\boldsymbol{q}\nu_{}}$ in Eqn.~\ref{eqn_coherent} is phonon linewidth ($\Gamma_{\boldsymbol{q}\nu_{}} = 1/\tau_{\boldsymbol{q}\nu_{}}$) and $V^{\alpha}_{\boldsymbol{q}, \nu\nu_{1}}$ is the $\alpha$-component of velocity operator obtained as:
\begin{equation}
    {\boldsymbol{V}}_{\boldsymbol{q}, \nu\nu_{1}} = 
    \frac{1}{2\sqrt{ \omega_{\boldsymbol{q}\nu_{}} \omega_{\boldsymbol{q}\nu_{1}} }} 
    \bra{ {{{\boldsymbol{e}}}_{\boldsymbol{q}\nu}}} 
    \frac{\partial \boldsymbol{D}_{\boldsymbol{q}} }{\partial \boldsymbol{q}} 
    \ket{ {{{\boldsymbol{e}}}_{\boldsymbol{q}\nu_{1}}}}.
\end{equation}

}

The open-source quantum mechanical simulation package Quantum Espresso is used to perform the DFT calculations with scalar-relativistic PBE exchange-correlation functional based ultra-soft pseudopotentials. The planewave kinetic energy cutoff and electronic wavevector grid are set at 60 Ry and 2$\times$2$\times$2 respectively. The total energy change is less than $5\times10^{-4}$ Ry/atom  on increasing the kinetic energy cutoff to 100 Ry or electronic wavevector grid to 6$\times$6$\times$6. The structures are fully relaxed to ensure residual forces are less than 10$^{-5}$ Ry/Bohr.  The relaxed lattice constants as obtained using these settings are 10.95 {\AA}  and 10.86 {\AA} for BGG and SGG respectively which are an over-prediction of experimentally
measured lattice constants by 2\% and 4\% \cite{eisenmann1986} and is a known shortcoming of PBE functionals \cite{Harl2009}.

The harmonic force constants are initially obtained on a 2$\times$2$\times$2 phonon wavevector grid using the DFPT calculations and are later interpolated to a 8$\times$8$\times$8 grid during the renormalization process. The cubic and quartic force constants are obtained by force-displacement data fitting on a 2$\times$2$\times$2 computational cell consisting of a total of 432 atoms. 150 such thermally displaced cells are used in this study resulting in a total of 194400 equations. The cubic force constant interactions are included up to  5 {\AA} and the quartic force constant interactions are limited to first neighbor shell. {\color{black}The three-phonon scattering rates are obtained on a phonon wavevector grid of size $8\times8\times8$. The obtained total thermal conductivities are converged to within 5\%  with these choices of simulation parameters.} 

\section{Structure}
\label{sec_struct}

The crystal structure of type-I inorganic clathrate is shown in Fig.~\ref{fig_structure}. The structure belongs to cubic space group ${Pm}\overline{3}{n}$ with spacegroup number 223. The cage atoms occupy sites 6$c$, 16$i$, and 24$k$ and encapsulate the guest atoms located at sites 2$a$ and 6$d$. The primitive unit cell is composed of 54 lattice sites of which 46  are occupied by Ga (16$i$) and Ge (6$c$ and 24$k$) atoms and 8 (2$a$ and 6$d$) are occupied by Ba/Sr atoms. The cages at 2$a$ sites are isotropic and are composed of 12 regular pentagonal surfaces. The cages at 6$d$ sites are asymmetric with 12 pentagonal and 2 hexagonal surfaces. For 2$a$ cages, the bond stiffness corresponding to Ba and Sr guest atoms (calculated by displacing guest atom by $0.01$ $\text{\AA}$ in required directions) are 1.62 and 1.06 eV/$\text{\AA}^2$. For 6$d$ cages, the stiffness is anisotropic and is 0.34 (0.12) and 0.84 (0.46) eV/$\text{\AA}^2$ in directions parallel and perpendicular to six-membered rings for Ba (Sr) atom. 

Previous literature studies have suggested the off-centering of the guest atoms in 6$d$ cages for Sr atoms \cite{Chakaumakos2000,Christensen2006A,chulholee2008,Takasu2006}. There have been studies where theoretical models have been used to explain guest off-centering\cite{Bridges2004, Nakayama2011}, however, to the best of our knowledge, these studies are all fitted to explain the experimentally observed large thermal displacements of atoms and there are no  studies explaining guest off-centering using ab-initio methods. The attempts on using ab-initio calculations for SGG either resulted in imaginary modes with on-centered guest atom or the obtained guest frequencies are way lower than experimental measurements \cite{Madsen2005}. We believe that these anomalous theoretical  results
are  due to (\roml{1}) the use of finite-difference approach for the extraction of phonon frequencies, and (\roml{2}) non-accounting of anharmonic effects on phonon frequencies. In particular, the PES is relatively flat for Sr atoms (as is evident from very low stiffness of  $0.12$ eV/$\text{\AA}^2$) and  the obtained forces in finite-difference method are comparable to numerical noise, thus resulting in numerical displacement-size dependent frequencies in the findings of Christensen et al.\cite{Christensen2016}. Further, in the same study,  the authors obtained 1.6 times smaller phonon frequencies compared to that measured in experiments for low-lying optical modes using the experimentally extracted anisotropic harmonic force constants. 
To comply with experimental findings, authors assumed that somehow the atoms are experiencing isotropic forces around their equilibrium positions and the proposed isotropic model resulted in an excellent agreement with experimental results. As will be shown later, this under-prediction of frequencies from anisotropic harmonic model is due to ignorance of higher-order effects and with inclusion of anharmonicity, the obtained frequencies are in perfect agreement with experiments without the use of isotropic assumption (the later was assumed by authors due to random hopping of rattler atoms between different off-centered sites).

\begin{figure}
\centering
\includegraphics{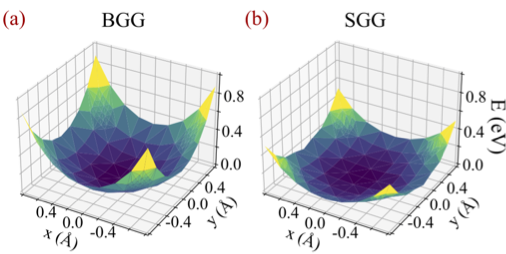}
\caption{
{The PES of (a) Ba and (b) Sr atoms around the 6$d$ centered site of tetrakaihedral cage in the plane parallel to the hexagonal face with cage center as the reference. The PES for Sr is  flatter compared to Br, thereby, indicating that Sr atoms are loosely bound and experience more anharmonic potential  as compared to Ba atoms.
}
}
\label{fig_PES}
\end{figure}

Our structure relaxation suggests energetically-stable on-centered positions for both Ba and Sr clathrates. We  explored the PES around the centered-positions by manually displacing rattler atoms in the plane parallel to hexagonal face of tetrakaidecahedral cage  and plot the resulting energies in Fig.~\ref{fig_PES}. We observe that for both clathrates, the on-centered position has lowest energy.
We have also carried out similar PES mapping for suggested off-centered site of Sr at ($0.6$, $0.6$, $0.1$) $\text{\AA}$ \cite{sales2001,Baumbach2005}  and found the similar behaviour. As such, we have used on-centered guest atoms for all calculations reported in this work.
This choice is further justified since the experimentally extracted energy barrier for off-centered tunneling of rattlers in SGG is $\sim$ 5 meV and is smaller than the thermal energy of atoms in the temperature range of interest (50-300 K) \cite{chulholee2008}.  It is worthwhile to emphasize that this 5 meV tunneling barrier is within the numerical noise of different simulation parameters used in this study and is, therefore, not resolvable here.

\section{Phonon Dispersions}
\label{sec_phondisp}

\begin{figure}
\begin{center}
\includegraphics{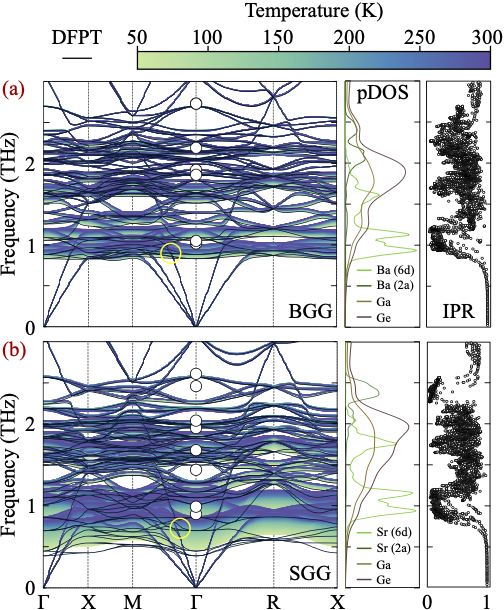}
\end{center}
\caption{The anharmonic phonon properties of (a) BGG and (b) SGG clathrates as obtained using the temperature-dependent sampling of the PES and quartic phonon renormalization. The open circles in dispersion plots represent experimentally measured Raman frequencies at a temperature of 300 K from Ref.~\cite{Takasu2006}. The density of states and inverse participation ratio are reported in side-figures for 300 K. The yellow circles in the $\Gamma-\text{M}$ direction in phonon dispersion plots highlight the  avoided-crossing of longitudinal acoustic phonons with low-lying optical phonons.}
\label{fig_harmonic}
\end{figure}

The anharmonic phonon dispersion of BGG and SGG clathrates as obtained using the temperature-dependent force constants are plotted in Fig.~\ref{fig_harmonic}(a) and \ref{fig_harmonic}(b) along with the atom-decomposed density of states and inverse participation ratio. The inverse participation ratio is a measure of mode localization and varies between $1/\text{N}_\text{unit}$ and 1 for completely localized and delocalized modes ($\text{N}_\text{unit}$ is the number of atoms in the unitcell, 54 for considered clathrates). 
{\color{black} The DFPT phonon dispersion in Fig. 3 corresponds to the lowest-order theory, i.e., under harmonic assumption and without considering T-dependent potential energy surface or phonon renormalization.} The experimentally measured Raman modes at 300 K are from Ref.~\cite{Takasu2006}

For both BGG and SGG, the acoustic phonons remain majorly unaffected with temperature and are delocalized with predominant contribution from cage atoms. In contrast, optical phonons undergo strong renormalization with temperature and become stiff at high temperatures. The effect is maximum for low frequency optical phonons and the frequencies become stiffer by 15 and 50\% for these phonons at the Gamma point at 300 K. These low frequency optical modes have predominant contribution from guest atoms centered inside $6d$ cages. Consistent with the rattling motion of these atoms, the inverse participation ratio of these modes is smaller than 1.

In agreement with the measurements of Christensen et al.~\cite{christensen2008}, avoided-crossing is obtained in the phonon dispersion of BGG  at $\boldsymbol{q}^{300K}_{ac}=(0.5\text{ }0.5\text{ }0.21)$ in the $\Gamma-\text{M}$ direction at 300 K. This avoided crossing between longitudinal acoustic and low-lying guest modes results in the flattening of heat-carrying acoustic modes and is argued as the reason for low thermal conductivity of filled clathrates. For SGG, the avoided-crossing occurs at $\boldsymbol{q}^{300K}_{ac}=(0.5\text{ }0.5\text{ }0.18)$ at a temperature of 300 K. With decreasing temperature, while the avoided-crossing moves closer to the Brillouin zone center for both BGG and SGG, the shift is much more pronounced in SGG [$\boldsymbol{q}^{50K}_{ac}=(0.5\text{ }0.5\text{ }0.18)$ and $(0.5\text{ }0.5\text{ }0.1)$ for BGG and SGG]. This suggests that (\roml{1}) at a temperature of 300 K,  Sr guest atoms result in a flattening of a larger fraction of heat-carrying acoustic cage phonons compared to Ba (and hence low thermal conductivity of SGG compared to BGG at 300 K) and (\roml{2}) the acoustic-phonon flattening becomes more pronounced at low temperatures; thereby suggesting a decrease in thermal conductivity with reducing temperature as is observed in experiments for SGG \cite{Nolas1998,Cohn1999,keppens2000,Chakaumakos2000,sales2001}. In the case of BGG, the temperature-dependence of avoided-crossing is much weaker and the temperature-dependence of the thermal conductivity is expected to follow the Umklapp-scattering and the thermal conductivity increases with reducing temperature \cite{keppens2000,sales2001,christensen2008}.

\section{Mean Square Displacements}
\label{sec_MSD}

\begin{figure}
\centering
\includegraphics{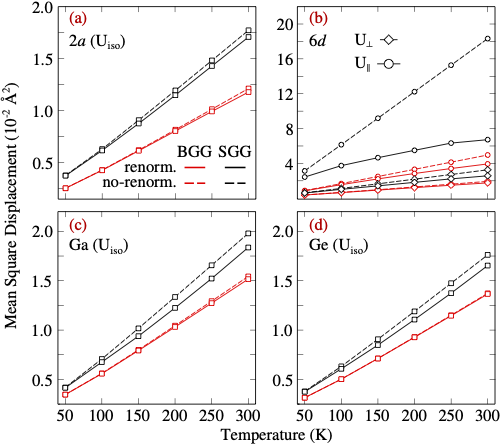}
\caption{
The mean squared thermal displacement of (a), (b) guest atoms centered at sites 2$a$ and 6$d$,  (c), (d) cage atoms Ga and Ge. The guest atoms centered at 6$d$ sites have anisotropic thermal displacements and $\text{U}_{\perp}$, $\text{U}_{\parallel}$ represent displacements perpendicular and parallel to the hexagonal face of the tetrakaidecahedra cage.
}
\label{fig_MSD}
\end{figure}

Moving further, the thermal mean square displacements are obtained using the temperature-dependent phonon dispersion for atoms sitting at different lattice sites and the results are presented in Fig.~\ref{fig_MSD}. 
For  guest atoms at 2$a$ sites and cage atoms, the thermal displacements are isotropic and varies between $0.01$-$0.02$ $\text{\AA}^2$ at 300 K which are consistent with experimental measurements \cite{Chakaumakos2000,Chakaumakos2001,sales2001,Qiu2004,Bentein2005,Christensen2006A,Christensen2009B} and are similar in range to other crystalline materials \cite{Sales1999A,Jain2020}. For guest atoms at 6$d$ sites, the displacements are anisotropic and are larger than $0.02$ $\text{\AA}^2$ in the plane parallel to  the hexagonal face of the tetrakaidecahedra cage. For BGG at 300 K, the obtained thermal displacements are $0.018$ and $0.039$ $\text{\AA}^2$ for 6$d$ cage atoms in the planes perpendicular and parallel to hexagonal face of tetrakaidecahedra cage. These numbers compare well with the experimental measured values of $0.015-0.018$ and $0.046-0.052$ $\text{\AA}^2$ by Christensen et al.\cite{Christensen2006A} For SGG, the corresponding calculated numbers are $0.025$ and $0.067$ $\text{\AA}^2$. The comparison with experiments is not possible for SGG due to a wide spread in the measured values ($0.03-0.08$ $\text{\AA}^2$ using the isotropic model and $0.03$ and $0.14$ $\text{\AA}^2$ using the anisotropic model\cite{Chakaumakos2000,Chakaumakos2001,sales2001,Qiu2004,Bentein2005}) .

The results reported in Fig.~\ref{fig_MSD} clearly suggest that the accuracy of lowest-order harmonic theory decreases at high temperatures. At a temperature of 300 K, the lowest-order theory  predicts erroneously large thermal displacement of $0.18$ $\text{\AA}^2$ for Sr atoms located at 6$d$ sites; thus highlighting its failure in describing anharmonic phonons and associated thermal properties of SGG.

\section{Lattice Thermal Conductivity}
\label{sec_thermalK}

\begin{figure}[h!]
\begin{center}
\includegraphics{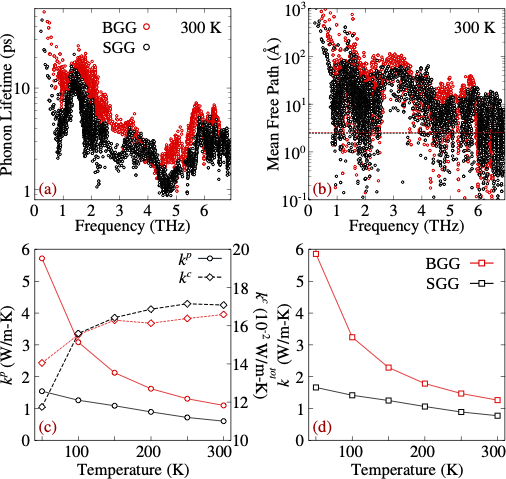}
\end{center}
\caption{{\color{black} (a) The mode-dependent phonon lifetimes and (b) phonon mean free paths for BGG and SGG clathrates as obtained by considering the three-phonon scattering processes at a temperature of 300 K. (c) The temperature-dependent contribution of phonon and coherent transport channels, and (d) the total thermal conductivity of BGG and SGG clathrates.
The dotted horizontal lines in (b) represent the Ioffe-Regel limit\cite{regel1960}.}}
\label{fig_thermalK}
\end{figure}

{\color{black} Next, we calculate the phonon scattering rates by considering the three-phonon processes and report the resulting mode-dependent lifetimes and mean free paths at a temperature of 300 K for BGG and SGG in Figs 5(a) and 5(b). It is recently reported that in rattler-like compounds\cite{}, the mean free paths of phonons could be shorter than the Ioffe-Regel limit and thereby question the validity of particle-like phonon picture \cite{regel1960,mukhopadhyay2018, simoncelli2019, Jain2020}. For considered clathrates, the obtained mean free paths for majority of the phonons are larger than the Ioffe-Regel limit. Nevertheless, we calculate the temperature-dependent thermal conductivities of both clathrates using the multi-channel transport model\cite{simoncelli2019} and report the results in Fig 5(c) and 5(d).
The contribution of coherent transport channel towards thermal transport is 13\% and 22\% in BGG and SGG at a temperature of 300 K. With reducing temperature, while the contribution of particle-like channel increase as 1/T, the contribution of coherent channel decrease, thereby resulting in a decrease in coherent contribution at lower temperatures. At a temperature of 300 K, our total predicted thermal conductivity of BGG is 1.26 W/m-K which is in close comparison with experimentally measured value of 1.31 W/m-K\cite{sales2001} and first-principles computations based value of 0.97 W/m-K\cite{Tadano2018} though the later was obtained without considering the contribution of coherent transport channel. For SGG, our predicted value is 0.77 W/m-K compared to an experimentally measured value of 1.01 W/m-K\cite{sales2001}. With reducing temperature, while the total predicted thermal conductivity increase for both BGG and SGG, the experimentally observed trend is crystal-like and glass-like for BGG and SGG respectively. We believe that these different temperature trends in the computed and experimentally measured thermal conductivity    of SGG could be due to (a) off-centering of guest atoms in experimentally synthesised samples of SGG\cite{Christensen2016}, (b) contribution from higher order phonon scattering and renormalization terms in the computed thermal conductivity of SGG, (c) electron-phonon interactions which alter the nature of thermal transport as reported earlier\cite{Avila2006,Bentien2004}}

\section{Conclusions}
\label{sec_conc}
To summarize, we obtained temperature-dependent phonon properties of type-I clathrates BGG and SGG accounting for anharmonic effects. In contrast with experiments, our calculations suggest stable on-center position for both Ba and Sr in tetrakaidecahedral cages which  could be a limitation of the pseudopotentials and/or simulation parameters used in this study. With on-centered guest atoms, we are able to reproduce experimentally observed  thermal displacements of rattler atoms at various temperatures. {\color{black} For thermal conductivity of BGG, we are able to reproduce the experimentally measured value at 300 K and the associated temperature-dependence. In case of SGG, while we find that the coherent channel contribution  to thermal transport is significant, the multi-channel thermal transport model with on-centering of guest atoms is insufficient in explaining the experimentally observed glass-like temperature-dependence of the thermal conductivity.}

\begin{acknowledgments} 
The authors acknowledge the financial support from IRCC-IIT Bombay and National Supercomputing Mission, Government of India (Grant Number: DST/NSM/R\&D-HPC-Applications/2021/10).  The calculations are carried out on SpaceTime-II supercomputing facility of IIT Bombay and PARAM Sanganak supercomputing facility of IIT Kanpur.
\end{acknowledgments}


\end{document}